\documentclass{acm_proc_article-sp}

\begin{document}

\title{Multiscale Topology of Chromatin Folding}

\numberofauthors{3}

\author{
\alignauthor
Kevin Emmett\\
       \affaddr{Departments of Physics and Systems Biology}\\
       \affaddr{Columbia University}\\
       \affaddr{New York, NY 10032}\\
       \email{kje2109@\\columbia.edu}
\alignauthor
Benjamin Schweinhart\\
       \affaddr{Center for Mathematical Sciences and Applications}\\
       \affaddr{Harvard University}\\
       \affaddr{Cambridge, MA 02138}\\
       \email{bschweinhart@\\cmsa.fas.harvard.edu}
\alignauthor
Raul Rabadan\\
       \affaddr{Departments of Biomedical Informatics and Systems Biology}\\
       \affaddr{Columbia University}\\
       \affaddr{New York, NY 10032}\\
       \email{rr2579@\\cumc.columbia.edu}
}

\maketitle

\begin{abstract}
The three dimensional structure of DNA in the nucleus (chromatin) plays an important role in many cellular processes.
Recent experimental advances have led to high-throughput methods of capturing information about chromatin conformation on genome-wide scales.
New models are needed to quantitatively interpret this data at a global scale.
Here we introduce the use of tools from topological data analysis to study chromatin conformation.
We use persistent homology to identify and characterize conserved loops and voids in contact map data and identify scales of interaction.
We demonstrate the utility of the approach on simulated data and then look data from both a bacterial genome and a human cell line.
We identify substantial multiscale topology in these datasets.
\end{abstract}



\section{Introduction}
\label{sec:introduction}
The 6 billion bases in the human genome would span a length of almost two meters if stretched end to end, yet occupy a compacted volume inside the nucleus of only a few $\mu\mathrm{m}^3$.
Even more remarkably, this million-fold level of compression is not random, but exhibits a complex hierarchical structure that intimately effects genome function through regulation of gene expression.
This multiscale pattern ranges from nucleosomes every 150 bases, promoter interactions at the megabase scale, topologically associated domains at the 10 megabase scale, and finally to organization of discrete chromosomes \cite{Dekker:2013hi}.
Chromatin conformation is dynamic, and will change throughout the cellular cycle, under the infuence of a diverse range of chromatin remodeling proteins, such as CTCF.
Chromatin architecture can further be controlled epigenetically through post-translational modifications including methylation and phosphorylation.

Recently developed experimental approaches have provided unprecedented high-throughput access into the three dimensional architecture of DNA inside the nucleus \cite{LiebermanAiden:2009jz,Dekker:2013hi,Ay:2015gv}.
These methods, known as \emph{chromosome conformation capture} (3C), use next-generation sequencing to probe for enriched physical proximity between nonadjacent genomic loci.
Hi-C couples 3C with ultra-deep sequencing to measure genome-wide interaction patterns in an unbiased manner.
However, while chromatin may fold in three dimensions, Hi-C contact data is only an indirect representation of these spatial relationships.
Several approaches have been developed to use contact map information to generate 3D embeddings of chromatin, however this introduces additional uncertainty in the analysis \cite{Ay:2015gv}.
Further, the contact map is an average over an ensemble of configurations.
We would therefore like to directly characterize topological properties of the ensemble without the need for such an embedding.

Topological data analysis (TDA) has been applied to several problems in genomics \cite{Chan:2013,Emmett:2014a}.
In this brief note we introduce the use of TDA to characterize the complex structure of chromatin inside the nucleus.
Our primary tool is persistent homology, which extracts global information about geometric and topological invariants in data.

We first demonstrate the approach on data from simulated polymer folding.
We then consider data from \emph{C. crescentus}, a circular bacterial genome.
Finally, we apply our approach to human cell line data, showing how persistent homology can capture complex multiscale folding patterns.
As we show, tools from topological data analysis may prove powerful at analyzing chromatin interaction data.

\section{Background}
\label{sec:background}
Hi-C contact data is generated as follows:
First, DNA is cross-linked in formaldehyde, linking segments of chromatin that are close in spatial proximity.
This step links pieces of chromatin that are in spatial proximity.
Second, pieces are fragmented and ligated to form closed loops.
Finally, pieces are sheared and sequenced, and the ends of each read are mapped to loci on the genome.
The data is summarized as a contact map representing counts of interactions between nonadjacent loci.
For more details, see \cite{Dekker:2013hi}.
From raw frequency data, normalization procedures are then applied to account for biases.
Normalization is a difficult problem and several such methods have been developed, see \cite{Ay:2015gv} for discussion.
In this work we largely use normalized contact matrices as input.
Pearson correlation $\rho$ measures similarity between loci, which we convert to a distance as $d=1-\rho$.

\begin{figure}
       \centering
       \includegraphics[width=\columnwidth]{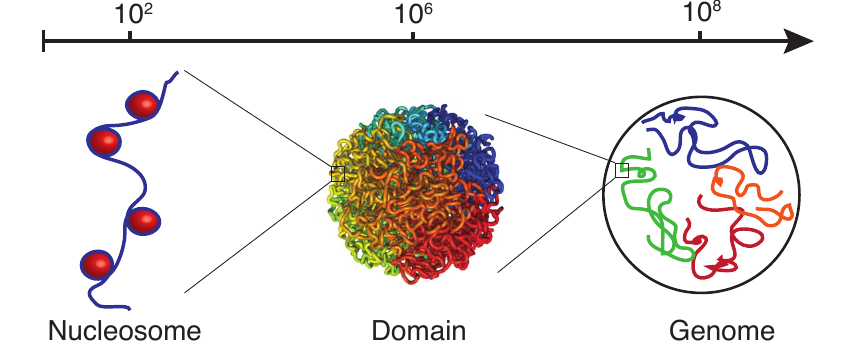}
       \caption{Three hierarchices of chromatin organization. At the 100 bp scale, DNA chains wrap around protein complexes called nucleosomes. At the megabase scale, these chains are compacted into domains. Lieberman-Aiden \emph{et al.} proposed closed domains form a \emph{fractal globule} structure. At the genome scale, chromosomes fold into the nucleus in separate territories. The fractal globule represents densely packed regions not open for transcription. Fractal globule image from \cite{LiebermanAiden:2009jz}. Reprinted with permission from AAAS.}
       \label{fig:chromatin_cartoon}
\end{figure}

Hi-C experiments have identified topologically associated domains.
Existing computational analyses have focused on identifying significant off-diagonal contacts and associating them with specific genomic interactions.
Here we focus on the global scales of chromatin folding.

We use persistent homology to analyze Hi-C contact maps.
Persistent homology captures information about loops and voids in a dataset using homology.
Homology information is tracked across a scale parameter $\epsilon$ via a series of nested simplicial complexes (see \cite{Carlsson:2014cn} for more details).
Invariants are summarized in a barcode diagram indexed by homology dimension $H_d$.
Each bar in the diagram, indexed as $PH_{i}$, is annotated with a birth time, $b_i$, and a death time, $d_i$.
$H_1$ gives information about looping between loci, and $H_2$ gives information about voids.
Following \cite{MacPherson:2012eq}, we define the \emph{size} of a PH class as

\begin{equation}
x_i = \frac{b_i+d_i}{2}.
\end{equation}

The distribution of PH class sizes reflects the scales of folding observed.
We use Dionysus to compute persistent homology \cite{Morozov:2012}.

\subsection{Long-Range Chromatin Interactions}
\begin{figure}
       \centering
       \includegraphics[width=\columnwidth]{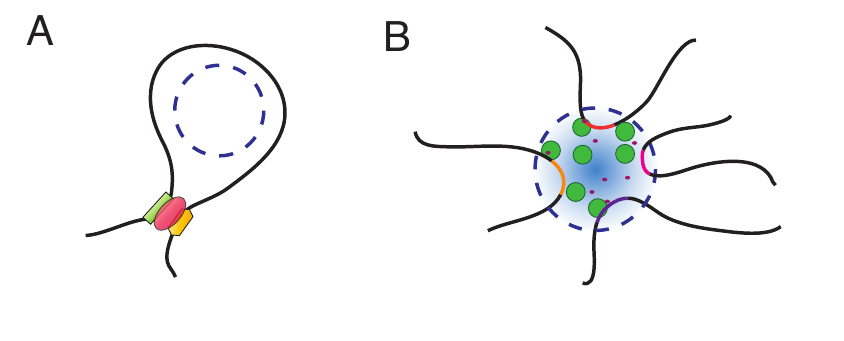}
       \caption{Two examples of long range chromatin interactions resulting in a topological loop. (A) A protein mediated (red) point-interaction between an enhancer (green) and promoter (yellow) sequence. (B) A transcription factory consists of dense RNA polymerase (green) around a structural core in which adjacent genomic loci (colored segments) will be cotranscribed. Transcription factors (purple) are shown.}
       \label{fig:chromatin_interactions}
\end{figure}

Long-range chromatin interactions can manifest in a number of different biological consequences at megabase scales.
In Figure~\ref{fig:chromatin_interactions} we show a cartoon of two possible types of interaction.
On the left we see a single interaction mediated by a binding protein that brings two nonadjacent loci into contact.
This could reflect a promoter-enhancer interaction, for example.
We call this interaction a \emph{one-jump loop}.
On the right, we see a more complex interaction representing multiple nonadjacent loci surrounding a dense compartment of polymerase proteins.
This phenomenon is known as a transcription factory, and genes adjacent to a given factory will have correlated levels of expression.
We call this interaction a \emph{multi-jump loop}, because the minimal hole generated by the filtration will span multiple nonadajacent loci.

\subsection{Minimal Cycle Algorithm}
\label{subsec:minimal_cycles}
It is important to be able to localize a cycle in order to annotate particular loops an interactions.
To define a notion of minimal cycle corresponding to a PH class, we first use the contact map to locate an ``essential edge'' which a cycle must contain.
To do this, the values of the heatmap are perturbed so that they are unique and there is a well-defined map from PH birth times to pairs of chromatin segments.
That is, we can associate to each PH class $(b_i, d_i)$ a unique ``essential edge'' that enters the filtration at the birth time $b_i$.
We define a minimal cycle corresponding to $(b_i,d_i)$ to be one containing the essential edge that traverses the shortest length along the genome, and is homologically independent from the minimal cycles of all classes born before $b_i$ \cite{Schweinhart:2015wr}.
This does not uniquely specify a cycle, and we break ties by preferring cycles with shorter jumps. 
Specifically, if $x~y~...~z$ is homologous to $x~x+1~y~...~z$ (where $x>y$) then the latter is considered better.
We use a breadth-first search starting with the essential edge to locate a minimal cycle for a given PH class.
Then, we shorten any jumps if possible.

\section{Polymer Simulations}
\label{sec:polymer_simulations}
To explore the use of topological methods for analyzing chromatin data, we used code from \cite{Doyle:2014ct} to simulate equilibrium folded polymer conformations.
The model uses a Monte Carlo approach to simulate chromatin as a one-dimensional polymer chain confined to a volume and allowed to come to an equilibrium conformation.
After equilibration, the 3D distance between monomers can be used as a measure of the contact frequency.

\begin{figure}
       \centering
       \includegraphics[width=\columnwidth]{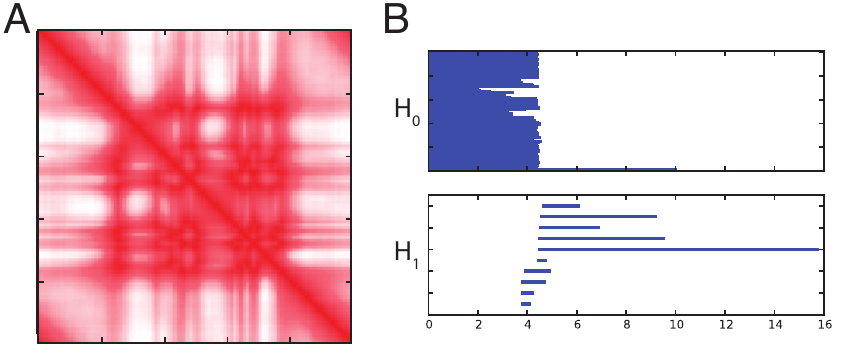}
       \caption{Polymer Simulation. (A) 50 Mb polymer with 10 fixed loops is allowed to reach an equilbrium conformation. (B) PH identifies 10 $H_1$ loops.}
       \label{fig:polymer_sim}
\end{figure}

In Figure~\ref{fig:polymer_sim} we show the output of one such a simulation.
Here, we simulated a 50 megabase chromatin segment as a chain of 1,000 15 nm monomers.
Each monomer corresponds to approximately 6 nucleosomes, or 1200 bp.
We inserted 10 fixed loops into the chain at random positions on the interior of the chain.
These loops represent recurrent protein-mediated interactions and mimic chromatin folding patterns observed in real data.
An ensemble of 5,000 conformations were generated and then averaged to yield the contact map depicted on the left.
On the right, we show the output of persistent homology on the average contact map.
Persistent homology recovers 10 $H_1$ intervals, consistent with the simulation and showing that topological information can be extracted from Hi-C-like contact maps.

\section{Caulobacter Data}
\label{sec:caulobacter_data}
We examined interaction data from \emph{Caulobacter crescentus} as published in \cite{Le:2013ci}.
\emph{C. crescentus} has a 4MB circular genome.
In that paper, chromatin interaction domains (CIDs) were identified at scales between 30 to 420 kb.
The authors proposed a structural model consisting of brush-like plectonemes arranged along the circular fiber.

\begin{figure}[ht]
       \centering
       \includegraphics[width=\columnwidth]{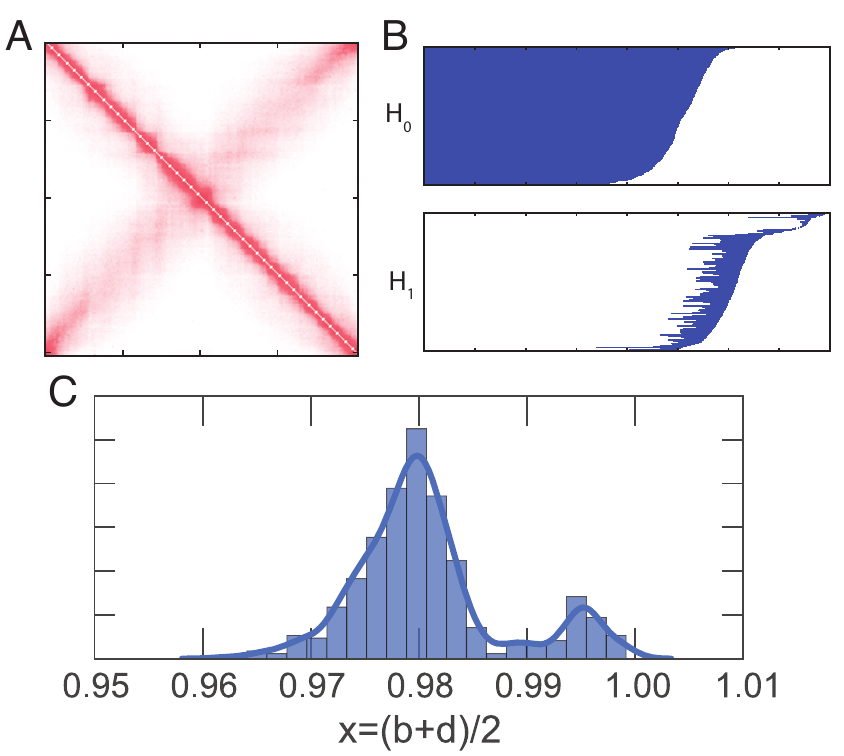}
       \caption{(A) Contact map and (B) barcode diagram for \emph{Caulobacter}. (C) Distribution of $H_1$ bar sizes for \emph{Caulobacter} shows a bimodal scale of folding patterns.}
       \label{fig:caulobacter_v2}
\end{figure}

Here we look at look at sample GSM1120446, a wildtype \emph{Caulobacter} cell.
In Figure~\ref{fig:caulobacter_v2}A we show the contact map data binned at 10 kb resolution.
Clearly identifiable are the strong interactions along the diagonal, as well as the circular off-diagonal interactions.
In Figure~\ref{fig:caulobacter_v2}B is the barcode diagram computed from this contact map.
Finally, in Figure~\ref{fig:caulobacter_v2}C we see that the size of $H_1$ invariants is strongly bimodally distributed.

\begin{figure}
       \centering
       \includegraphics[width=\columnwidth]{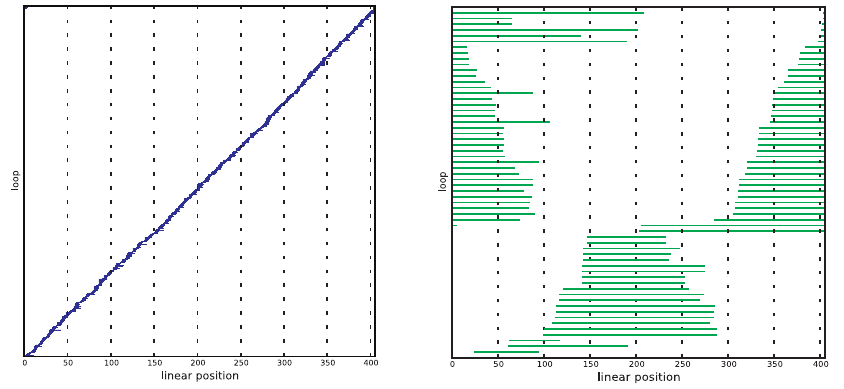}
       \caption{Minimal cycles projected linearly for \emph{Caulobacter}. Left: small loops distribute uniformly across the genome. Right: pattern of large loops, which segregate into two chromosomal domains.}
       \label{fig:caulobacter_bar_sizes}
\end{figure}

We used the minimal-cycle algorithm to determine a representative basis for each $H_1$ loop.
Figure~\ref{fig:caulobacter_bar_sizes} shows the set of minimal cycles arranged along the genomic axis.
We divide the loops between small- and large-scale loops as identified in Figure~\ref{fig:caulobacter_v2}C.
On the left, we see that the small-scale loops cover small genomic scales and are regularly distributed along the genome.
These small loops may associate to small nucleoid-associated proteins or structural maintenance complexes (see \cite{Wang:2013it}), or may simply reflect stochastic folding.
On the right, we see the large-scale loops cover broader genomic regions (average size 100kb).
These loops do not associate with the CIDs identified in \cite{Le:2013ci}, but rather reflect larger scale folding patterns.



\section{Human Data}
\label{sec:human_data}
We examined one of the original human Hi-C data sets as published in \cite{LiebermanAiden:2009jz}.
In that paper, the authors proposed a two-compartment model of chromosomal organization associated with open and closed chromatin states and correlated with gene expression patterns.
At the megabase scale, they proposed a fractal globule model in which nearby loci along the polymer are spatially proximate in 3D.

\begin{figure}
       \centering
       \includegraphics[width=\columnwidth]{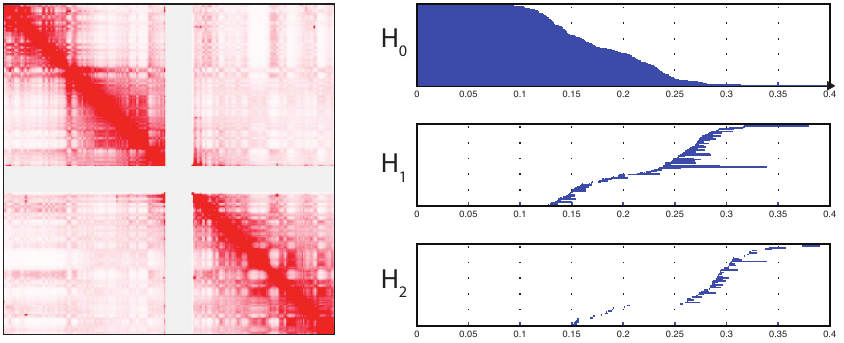}
       \caption{Hi-C data chromosome 1 from GM06690 human cell line data, from \cite{LiebermanAiden:2009jz}. Left: Contact map representation. Right: PH identifies complex multiscale topology.}
       \label{fig:human_data}
\end{figure}

We looked at data from GM06690, a healthy human lymphoblastoid cell line.
In Figure~\ref{fig:human_data} we show an example from chromosome 1 measured at 1 MB resolution.
On the left is the observed contact map.
The gray band in the middle represents the position of the centromeres.
On the right is the barcode diagram computed persistent homology.
We observe substantial structure in both $H_1$ and $H_2$.

\begin{figure}
       \centering
       \includegraphics[width=\columnwidth]{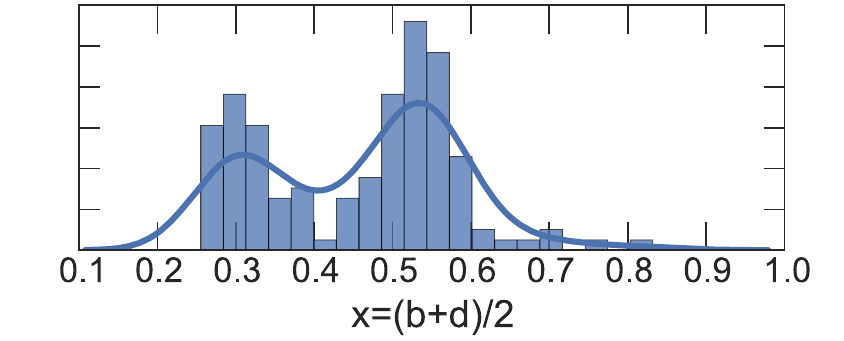}
       \caption{Distribution of $H_1$ bar sizes for GM06690 human cell line data shows a bimodal scale of folding patterns.}
       \label{fig:human_data_histogram}
\end{figure}

In Figure~\ref{fig:human_data_histogram} we show the distribution of $H_1$ bar sizes.
We observe a strong bimodal structure, representative of two scales of chromatin folding.
This is consistent with the results in \cite{LiebermanAiden:2009jz}, which identified topologically associated domains at the 10MB scale.

Because the contact map is at 1 MB resolution, it is too coarse to capture nucleosome-level folding patterns (200bp).
More recent work has yielded Hi-C datasets at kilobase resolution \cite{Jin:2013hm,Rao:2014eo}, however at this resolution the contact map is too large for an efficient persistent homology computation across the entire genome (or even an entire chromosome).
It is possible that the linear nature of the chain may make a heuristic homology computation feasible.

\section{Conclusions}
\label{sec:conclusions}
Patterns of chromatin conformation inside the nucleus exhibit complex, multiscale structures that are intimately tied to genome function.
Here we have used methods from TDA to characterize the scale and conformation of these structures.
TDA is a natural framework to study data of this type because there is a clear definition of the ambient embedding space.
Using simulation we showed that persistent homology captures recurrent loops.
In real data, we observed multiscale structures reflecting hierarchical patterns of chromatin organization.
In the present work we have examined only intrachromosomal interactions.
Interchromsomal interactions have also been reported, however the resulting contact maps are too large for homology computations at sufficient resolution.
Future work will focus on identifying heuristics to improve these calculations.

\section{Acknowledgments}
KE and RR are supported by NIH grants U54-CA193313-01 and R01-GM117591-01.
BS is supported by the Center of Mathematical Sciences and Applications at Harvard University.
The authors thank Robert MacPherson, Arnold Levine, and Nils Baas for useful discussions.

\bibliographystyle{abbrv}
\bibliography{topdrim4bio-HiC}

\begin{thebibliography}{10}

\bibitem{Ay:2015gv}
F.~Ay and W.~S. Noble.
\newblock {Analysis methods for studying the 3D architecture of the genome}.
\newblock {\em Genome Biology}, 16(1):1306, Sept. 2015.

\bibitem{Carlsson:2014cn}
G.~Carlsson.
\newblock {Topological pattern recognition for point cloud data}.
\newblock {\em Acta Numerica}, 23:289--368, May 2014.

\bibitem{Chan:2013}
J.~Chan, G.~Carlsson, and R.~Rabadan.
\newblock Topology of viral evolution.
\newblock {\em Proceedings of the National Academy of Sciences},
  110(46):18566--–18571, 11 2013.

\bibitem{Dekker:2013hi}
J.~Dekker, M.~A. Marti-Renom, and L.~A. Mirny.
\newblock {Exploring the three-dimensional organization of genomes:
  interpreting chromatin interaction data}.
\newblock {\em Nature Reviews Genetics}, 14:390--403, 2013.

\bibitem{Doyle:2014ct}
B.~Doyle, G.~Fudenberg, M.~Imakaev, and L.~A. Mirny.
\newblock {Chromatin Loops as Allosteric Modulators of Enhancer-Promoter
  Interactions}.
\newblock {\em PLoS Computational Biology}, 10(10):e1003867, Oct. 2014.

\bibitem{Emmett:2014a}
K.~J. Emmett and R.~Rabadan.
\newblock Characterizing scales of genetic recombination and antibiotic
  resistance in pathogenic bacteria using topological data analysis.
\newblock In D.~Slezak, A.-H. Tan, J.~F. Peters, and L.~Schwabe, editors, {\em
  Brain Informatics and Health}, volume 8609 of {\em Lecture Notes in Computer
  Science}, pages 540--551. Springer, 2014.

\bibitem{Jin:2013hm}
F.~Jin, Y.~Li, J.~R. Dixon, S.~Selvaraj, Z.~Ye, A.~Y. Lee, C.-A. Yen, A.~D.
  Schmitt, C.~A. Espinoza, and B.~Ren.
\newblock {A high-resolution map of the three-dimensional chromatin interactome
  in human cells}.
\newblock {\em Nature}, Oct. 2013.

\bibitem{Le:2013ci}
T.~B. Le, M.~V. Imakaev, L.~A. Mirny, and M.~T. Laub.
\newblock {High-resolution mapping of the spatial organization of a bacterial
  chromosome}.
\newblock {\em Science}, 342(6159):731--734, 2013.

\bibitem{LiebermanAiden:2009jz}
E.~Lieberman-Aiden, N.~L. van Berkum, L.~Williams, M.~Imakaev, T.~Ragoczy,
  A.~Telling, I.~Amit, B.~R. Lajoie, P.~J. Sabo, M.~O. Dorschner, R.~Sandstrom,
  B.~Bernstein, M.~A. Bender, M.~Groudine, A.~Gnirke, J.~Stamatoyannopoulos,
  L.~A. Mirny, E.~S. Lander, and J.~Dekker.
\newblock {Comprehensive Mapping of Long-Range Interactions Reveals Folding
  Principles of the Human Genome}.
\newblock {\em Science}, 326(5950):289--293, Oct. 2009.

\bibitem{MacPherson:2012eq}
R.~MacPherson and B.~Schweinhart.
\newblock {Measuring shape with topology}.
\newblock {\em Journal of Mathematical Physics}, 53(7):073516, 2012.

\bibitem{Morozov:2012}
D.~Morozov.
\newblock Dionysus: a {C++} library for computing persistent homology, 2012.

\bibitem{Rao:2014eo}
S.~S.~P. Rao, M.~H. Huntley, N.~C. Durand, E.~K. Stamenova, I.~D. Bochkov,
  J.~T. Robinson, A.~L. Sanborn, I.~Machol, A.~D. Omer, E.~S. Lander, and E.~L.
  Aiden.
\newblock {A 3D Map of the Human Genome at Kilobase Resolution Reveals
  Principles of Chromatin Looping}.
\newblock {\em Cell}, pages 1--16, Dec. 2014.

\bibitem{Schweinhart:2015wr}
B.~Schweinhart.
\newblock {\em {Statistical Topology of Embedded Graphs}}.
\newblock PhD thesis, Princeton University Press, July 2015.

\bibitem{Wang:2013it}
X.~Wang, P.~M. Llopis, and D.~Z. Rudner.
\newblock {Organization and segregation of bacterial chromosomes}.
\newblock {\em Nature Reviews Genetics}, 14(3):191--203, Feb. 2013.

\end{thebibliography}

\end{document}